\documentclass{emulateapj}

\def\gax{\mathrel{\raise.3ex\hbox{$>$}\mkern-14mu\lower0.6ex\hbox{$\sim$}}}
\def\lax{\mathrel{\raise.3ex\hbox{$<$}\mkern-14mu\lower0.6ex\hbox{$\sim$}}}
\def\gtorder{\mathrel{\raise.3ex\hbox{$>$}\mkern-14mu
             \lower0.6ex\hbox{$\sim$}}}
\def\ltorder{\mathrel{\raise.3ex\hbox{$<$}\mkern-14mu
             \lower0.6ex\hbox{$\sim$}}}

\begin{document}

\title{The Mid-IR Contribution Of Dust Enshrouded Stars In Six Nearby Galaxies}

\author{J.R. Gerke$^{1}$, C.S. Kochanek$^{1,2}$}
 
\altaffiltext{1}{Department of Astronomy, The Ohio State University, 140 West 18th Avenue, Columbus OH 43210.}
\altaffiltext{2}{Center for Cosmology and AstroParticle Physics, The Ohio State University, 191 W. Woodruff Avenue, Columbus OH 43210.}

\begin{abstract}
We measure the integrated contributions of dusty AGB stars and other luminous red mid-IR sources to the mid-IR luminosities of 6 galaxies (M81, NGC~2403, NGC~300, M33 and the Magellanic Clouds).  We find the dusty AGB stars whose mid-IR fluxes are dominated by dust rather than photospheric emission contribute from 0.6\% (M81) to 5.6\% (SMC) of the $3.6~\micron$ flux and 1.0\% (M81) to 10.1\% (SMC) of the $4.5~\micron$ flux.  We find a trend of decreasing AGB contribution with increasing galaxy metallicity, luminosity and mass and decreasing SSFR.  However, these galaxy properties are strongly correlated in our sample and the simplest explanation of the trend is galaxy metallicity.  
 Bright, red sources other than dusty AGB stars represent a smaller fraction of the luminosity, $\sim$1.2\% at 3.6~$\micron$, however their dust is likely cooler and their contributions are likely larger at longer wavelengths. 
Excluding the SMC, the contribution from these red sources correlates with the specific star formation rate as we would expect for massive stars.  In total, after correcting for dust emission at other wavelengths, the dust around AGB stars radiates 0.1-0.8\% of the bolometric luminosities of the galaxies.  Thus, hot dust emission from AGB and other luminous dusty stars represent a small fraction of the total luminosities of the galaxies but a significant fraction of their mid-IR emissions.

\end{abstract}


\section{Introduction}

Understanding the emissions from AGB stars is important for stellar evolution, models of galaxy spectral energy distributions (SEDs) and inferences about stellar populations and their evolution.  AGB stars are H and He shell burning stars that evolve from ~0.8-8 $M_{\odot}$ Main Sequence stars.  During the AGB phase, which lasts 1-13 Myr \citep{vass1993}, the star can undergo significant mass-loss, particularly in the Thermal-Pulsing AGB phase (TP-AGB, $\sim1.7M_{\odot} \le M \le 6M_{\odot}$).  During this last 0.2-2 Myr of the AGB phase, the AGB star experiences thermal pulses from a series of shell flashes that drive high mass loss rate winds conducive to the formation of dust.
The short lifetime of this phase makes these stars relatively rare, so they are generally absent from the star clusters used to calibrate stellar evolution models.  As a result, this is one of the most uncertain phases of stellar evolution.
Their large range of mass loss rates combined with dust formation mean that AGB stars can be important over a broad range of wavelengths from the optical to the mid-IR. 

For distant galaxies, the properties of stellar populations must be inferred from spectra or the SEDs of ensembles of stars.  Since AGB stars represent a non-trivial fraction of the luminosity of $\sim$Gyr old stellar populations, any uncertainties in the AGB phase propagate into uncertainties in overall galaxy properties because the stellar population synthesis (SPS) model must assume some treatment of the (TP) AGB stars. \citet{maraston2006} shows how different calibrations of the TP-AGB stars in the SPS models can cause large, systematic differences in the fits to SEDs and the resulting estimates of a galaxy's mass and age.  \citet{conroy2010} and \citet{conroy2010b} show in a more general way how TP-AGB stars can effect the determination of a galaxy's properties. 
 
Much progress has been made recently in moving beyond the Milky Way and Magellanic Cloud star clusters to calibrate the AGB phase of stellar evolution (e.g. \citealt{kriek2010}; \citealt{meidt2012b}; \citealt{zibetti2012}).  In particular, \citet{melbourne2012} used NIR observations to build on the optical study of 12 metal poor galaxies by \citet{girardi2010} to constrain the lifetimes of AGB stars based on the number and luminosities of the AGB stars. While they found that the numbers of observed AGB stars agreed with the updated \citep{girardi2010} SPS models given their uncertainties, the predicted luminosities are too large. 
They go on to demonstrate how more accurate lifetimes and luminosities are needed to be able to accurately describe galaxies at high redshift.  \citet{boyer2009} examined AGB stars in eight Local Group dwarf irregular galaxies in the Spitzer IRAC bands.  They examined the number of AGB stars, their mass-loss rates and how this mass returning to the ISM could effect the current SFR.  They also suggest that optical AGB searches will miss 30\%-40\% of AGB stars due to dust obscuration.  

In this paper we focus on the contribution of hot dust around luminous stars, particularly AGB stars, to the mid-IR emission of 6 nearby galaxies of varying properties. In addition to AGB stars, these sources include dusty young star clusters and other luminous stars that have experienced dense, episodic winds or eruptive outbursts.  We use archival Spitzer IRAC 3.6 and $4.5~\micron$ observations of M33, M81, NGC~300, NGC~2403, the SMC and the LMC to survey these populations.
We are focusing on AGB stars whose mid-IR fluxes are dominated by dust rather than photospheric emission, and we will refer to them as DAGB (dusty AGB) stars in order to distinguish them from the remainder of the AGB population whose mid-IR emission is predominately photospheric. 
  Section 2 describes the data and its analysis.  We identified these stars as those with red [3.6]$-$[4.5] colors indicating the presence of hot circumstellar dust.  The same color criteria also identify other, more luminous stars whose mid-IR fluxes are dominated by hot dust, and we will examine these as a separate population of ``red'' stars.  Section 3 presents the data analysis, including the source classification and background corrections.  In Section 4 we determine the fraction of the [3.6] and [4.5] flux contributed by the DAGB  and red stars and examine the trends in the DAGB and red star flux fractions with galaxy properties.  Our final summary is in Section 5.

\section{Data and Analysis}
We used archival {\it Spitzer} IRAC \citep{fazio2004} 3.6 and 4.5~$\micron$ images of M81, NGC~300, NGC~2403 and M33.  For the SMC and LMC we used the SAGE catalogs (\citealt{gordon2011}; \citealt{meixner2006}).  The M33 data were the six co-added epochs from \citet{mcquinn2007} that were also reanalyzed by \citet{thompson2009}. The Local Volume Legacy (LVL) Survey \citep{dale2009} data were used for NGC~300. For M81 and NGC~2403, we used the SINGS Legacy Survey \citep{kennicutt2003} data.  The M33 data have a pixel scale of $1\farcs2$ pixel$^{-1}$, while NGC~300, M81 and NGC~2403 have a pixel scale of $0\farcs75$ pixel$^{-1}$.  We initially examined the [5.8] and [8.0] band data as well, but resolution-induced confusion made this problematic and we decided to exclude these bands from further analysis. 

\begin{figure}
\plotone{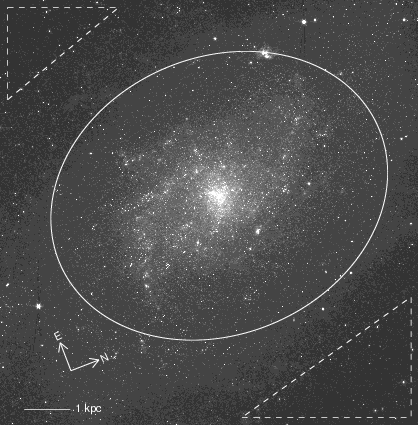}
\figurenum{1}
\label{fig:M33}
\caption{Spitzer 3.6$\micron$ image of M33. The solid curve shows the region used for the galaxy and the dashed lines outline the background regions.}
\end{figure}

For each mosaic we defined the galaxy as an ellipse centered at the same coordinates and with the same position angle and ellipticity as used by the LVL Survey.  The size of the ellipse was generally limited by the field of view of the observation.  For M33 we used a semi-major axis of 13\farcm5.  For NGC~300, M81 and NGC~2403 we used a semimajor axis of 10\farcm2.  For the SMC and LMC we used circles visually centered on the galaxies that were 2.0\degr~and 3.5\degr~in radius, respectively. We draw the sources and determine global fluxes for the galaxies from these areas. Figure~\ref{fig:M33} illustrates this for M33.

We identified sources on the IRAC mosaics using the DAOPHOT/ALLSTAR PSF fitting photometry package \citep{stetson1992}.  A minimum of 5 relatively isolated stars were used to determine the PSF, and sources were required to be at least 3$\sigma$ above the background at both [3.6] and [4.5] with a positional match of less than 1 pixel.  Stars with an apparent magnitude of [3.6] $<$ 8.5~mag were masked as foreground stars.  We chose this magnitude because it would include any possible $\eta$ Carinae in our nearest galaxy with images, M33, while removing the brightest foreground stars. We masked a 15 to 25 pixel radius, depending on the brightness of the masked star.  Using the distances in Table~\ref{tab:prop}, the fluxes were transformed to an absolute scale of $\nu L_{\nu}$ in solar luminosities. 

Table \ref{tab:prop} summarizes the properties of the galaxies.
Most of these are taken from the literature, as reported in the table caption. 
The star formation rates (SFR) were derived using H$\alpha$ measurements \citep{kennicutt2008} and calibrated to a SFR following \citet{meurer2009}.
The stellar masses for the four more distant galaxies were determined from the 2MASS \citep{skrutskie2006} K band magnitudes measured by \citet{dale2009}, assuming a mass-to-light ratio of 0.95 \citep{bell2003}.  The K-band mass-to-light ratio is less sensitive to star formation history and galaxy color \citep{bell2003} and will provide a good estimate for the galaxies in our sample.  Their metallicities are from \citet{zaritsky1994} and were calculated at half the semimajor axis of the galaxy.  
 While there is some controversy over the actual galaxy metallicities and gradients, it is the differential metallicities that are important. For example, \citet{urbaneja2005} present a central metallicity and gradient for NGC~300 that are much lower than is found in \citet{zaritsky1994}, however when they fit their data with the \citet{zaritsky1994} $R_{23}$ calibration, the metallicity estimates differ by only 0.05 at half the semimajor axis of the galaxy.  Such differences will not effect our conclusions.
The masses for the SMC and LMC were determined using the $3.6\micron$ flux from \citet{dale2009}, assuming a mass-to-light ratio found by taking the uncertainty weighted average of the 24 Sa galaxies from \citet{falcon2011}.  The metallicities of the SMC and LMC are taken from \citet{peimbert1976} and \citet{pagel1978}, respectively.  The specific star formation rate (SSFR) is the SFR per unit stellar mass. 

The total mid-IR luminosity of each galaxy was determined in two steps, with the measurements shown in Table~\ref{tab:totL}.  We first used the IRAF Ellipse task to measure the flux inside the previously defined elliptical regions.  To correct this flux for foreground and background emission, we next estimated the mean surface brightness in a background region outside the ellipse defining the galaxy and within the mosaic's area of equal exposure time. The background areas for M33, M81, NGC~300 and NGC~2403 are small sections located as far as possible from the galaxy given the mosaic geometry (see Figure \ref{fig:M33}).  The backgrounds for the LMC and SMC were determined from annuli with sections cut out where the exposure times differed from the central regions.  For the LMC we use an annulus from $3.7\degr$~to 4.0$\degr$ and for the SMC the annulus is from 2.3$\degr$ to 2.5$\degr$.  We then subtracted this mean background surface brightness multiplied by the area of the signal region from the flux found by the Ellipse task.  We have uniformly masked the rarer bright foreground stars, about 10 stars per image.  We cannot mask too many stars before starting to lose the brightest mid-IR sources in the target galaxies.  However, we tested masking a larger number of stars, and found only a modest effect on the luminosity estimates.
 Our background corrected galaxy luminosities are in good agreement \citet{dale2009}. The four more distant galaxies agree within 1$\sigma$, except for the 4.5~$\micron$ luminosities of M33 and NGC~300, which agree within 2$\sigma$. The LMC is the most discrepant, at $\sim4\sigma$ fainter.  This difference is likely caused by our smaller galaxy aperture.

\section{CMDs and Source Classification}
\begin{figure}
\plotone{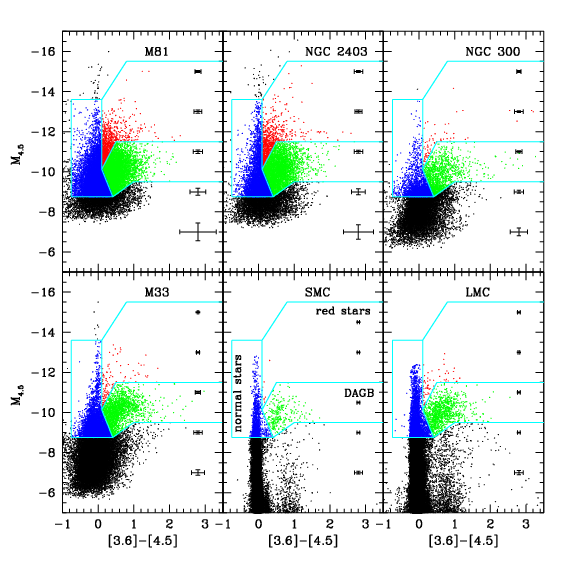}
\figurenum{2}
\label{fig:cmdall}
\caption{CMDs for all 6 galaxies.  The DAGB stars are in green, the luminous dusty stars are in red, and the normal stars are in blue.  The errorbars to the right indicate the typical photometric uncertainties as a function of magnitude.}
\end{figure}

We classified the sources using the $M_{4.5}$ and [3.6]$-$[4.5] color magnitude diagram (CMD).  Figure~\ref{fig:cmdall} shows the CMDs with classification criteria for all 6 galaxies.  We used a conservative lower luminosity limit of $M_{4.5}\le-8.75$ mag.  This limit is to ensure accurate photometry and matches between bands for all galaxies.  Color errors dominate any attempt to classify lower luminosity sources in the more distant galaxies. 
We divided the stars into three regions using the boundaries defined in Table \ref{tab:cmd}.  The boundaries were largely determined from the M33 CMD, which has a clear and well-populated DAGB sequence.
We first defined a region consisting of normal main sequence stars and evolved stars with little or no dust (blue).  In addition to the lower luminosity limit, we also used an upper limit to eliminate bright, foreground stars, a blue limit to eliminate peculiar outliers and a red limit to remove stars with hot dust. Next we defined the DAGB region (green) based on the CMD of M33.  Finally, we defined a region to include any other luminous dusty stars (red), which we will refer to as the red star region.  The upper luminosity limit was allowed to be higher here in order to include objects such as $\eta$ Carinae and dusty star clusters.

\begin{figure}
\plotone{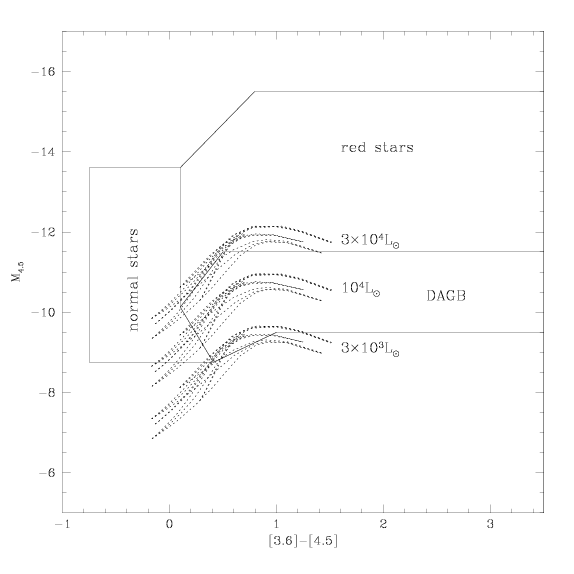}
\figurenum{3}
\label{fig:dust}
\caption{A CMD showing our empirical classification criteria based on the mid-IR CMD of M33 with solid lines. 
The dotted lines show models for AGB stars assuming various dust compositions from \citet{groenewegen2006}.  The models are scaled to luminosities of 3,000$L_{\odot}$, 10,000$L_{\odot}$ and 30,000$L_{\odot}$.
}
\end{figure}

Figure~\ref{fig:dust} superimposes the mid-IR properties of the DAGB models from \citet{groenewegen2006} on these empirical classifications based on the M33 mid-IR CMD.  The models for various dust properties are labeled with the luminosity of the star to which they have been scaled.  Our classification criteria for DAGB stars agree well with the colors and luminosities of these models. 

\begin{figure}
\plotone{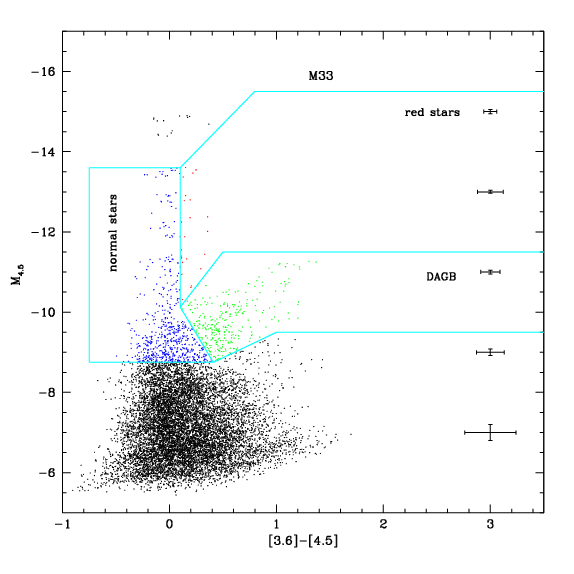}
\figurenum{4}
\label{fig:M33bg}
\caption{The CMD of the M33 background regions. To correct the density of the points for the smaller area of the background regions we have corrected the total number of objects for the area.  We randomly drew an area-corrected number of sources from the background region and applied a small random scatter to the magnitudes so that the background region CMD can be visually compared to Figure \ref{fig:cmdall}.}
\end{figure}

The total flux associated with each class of sources is then found by summing the fluxes of the individual stars.  There will also be contamination from both foreground stars and background extragalactic sources.  As with the total fluxes, we estimated this contamination using the background regions defined for each galaxy.  We built CMDs for each background region, computed the mean fluxes, and subtracted them from the total flux after adjusting for the differences in sampling area. 
 Table~\ref{tab:bg} shows the percentage of source light that is considered background for each of our source classes in each galaxy.  The background corrections can be quite large.  Figure~\ref{fig:M33bg} shows the background region CMD for M33 that has been corrected for area by randomly drawing sources from the background region and adding a small random scatter to the magnitudes so that the effective area is the same as for the galaxy CMD in Figure \ref{fig:cmdall}.  We can still see a weak DAGB feature in this background CMD, so the background region still includes stars in M33.  Ideally we would use background regions better separated from the galaxies to eliminate this problem, but we are limited by the observations and have chosen regions as far from the galaxies as the data allow.  While this implies some oversubtraction, we apply this background correction to both the summed luminosity of the sources and the estimates of the total galaxy luminosity.
Since we ultimately focus on the ratio of these two quantities, the final results should be little effected by any oversubtraction.  For example, the DAGB luminosity fraction determined for M33 using an elliptical background region extending from 17\farcm6 to 20\farcm0 rather than our standard region shown in Figure \ref{fig:M33}, changes the DAGB luminosity fraction by less than 0.1\% in both IRAC bands.  We carried out similar tests for each galaxy.  M33 showed the largest changes, and the differences are always smaller than our estimated statistical uncertainties.

We estimate the statistical uncertainties using bootstrap resampling.  For the total galaxy flux for M33, M81, NGC~300 and NGC~2403, we calculated the uncertainties in the background fluxes by bootstrap resampling the image in 5$\times$5 pixel squares.  Using patches comparable in size to point sources will roughly include the Poisson fluctuations of point sources in the uncertainties.  For the LMC and SMC, where we were working from catalogs, we bootstrap resampled the sources both from the background and galaxy regions.  We also bootstrap resampled the DAGB, red, and blue sources in both the galaxy and the background region to determine the uncertainties in both the total stellar flux for each category and their background contamination correction. 

Finally, we calculated the fraction of the total luminosity that is attributed to each type of star for both the original and the background-corrected values.  Table~\ref{tab:fracR} shows the fraction of the luminosity that the DAGB, red and normal stars contribute to each galaxy without any background correction, while Table~\ref{tab:fracC} shows these quantities corrected for background contamination.  We will discuss only the background corrected values.

\section{Results}
\begin{figure}
\plotone{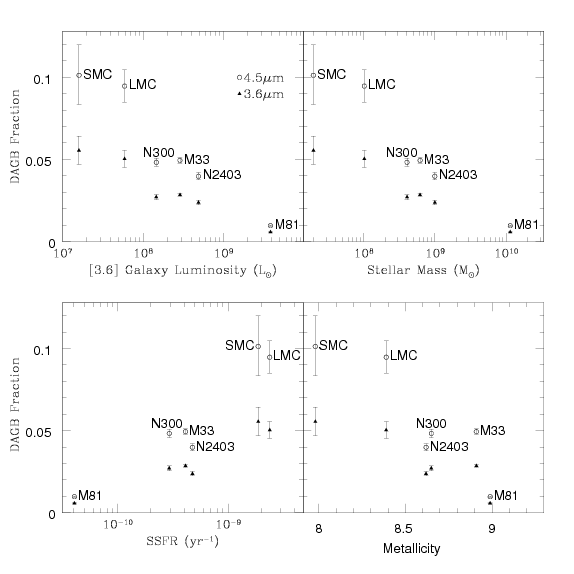}
\figurenum{5}
\label{fig:agbplot}
\caption{The fraction of the 3.6\micron~(filled triangles) and 4.5\micron~(open circles) luminosity produced by stars in the DAGB region as a function of galaxy luminosity (top left), stellar mass (top right), specific star formation rate (lower left) and metallicity (lower right). }
\end{figure}

We start by discussing the DAGB region, which accounts for 0.6\% (M81) to 5.6\% (SMC) of the $3.6~\micron$ flux and 1.0\% (M81) to 10.1\% (SMC) of the $4.5~\micron$ flux.  
Bear in mind that we are focusing on the DAGB stars whose 3.6 and $4.5~\micron$ luminosities are dominated by dust rather than photospheric emission.
As shown in Figure~\ref{fig:agbplot}, we see a general trend of decreasing DAGB contribution with increasing galaxy luminosity, mass and metallicity and an increasing DAGB light fraction with increasing SSFR. The most likely interpretation is that the DAGB fraction of the mid-IR luminosity is higher for the lower metallicity galaxies. 
While it appears the mid-IR flux contribution also correlates with luminosity, mass, and SSFR, we believe this is just a consequence of the lower metallicity galaxies having higher SSFRs and lower masses.  Since the DAGB stars are 0.5-2 Gyr old, the DAGB luminosity should have no intrinsic correlation with the current SFR or SSFR.  The star formation within that 1.5 Gyr window will be tightly correlated with the DAGB luminosity contribution (\citealt{kelson2010}; \citealt{kriek2010}), but we lack complete star formation histories for all the galaxies in our full sample and so cannot completely rule out SFR as the cause of the correlations.  Similarly, the stellar mass depends on the integrated SFR, which should also show no direct correlation.

To examine whether the results were sensitive to changes in the photometric depth with distance, we degraded the SMC and LMC photometry to match that for M81 and then recalculated the DAGB and other stellar fractions.  We found no significant changes to the results based on the true (smaller) uncertainties.  The changes observed were small and also in the direction of making the discussed trends stronger.  For example, the SMC  $3.6~\micron$ DAGB fraction changed from ($5.6\pm0.9$)\% to ($6.4\pm1.0$)\% after expanding the scatter.  The largest change was in the ``red stars'' fraction, due to small number statistics, but even this change was still within $1\sigma$. 

\begin{figure}
\plotone{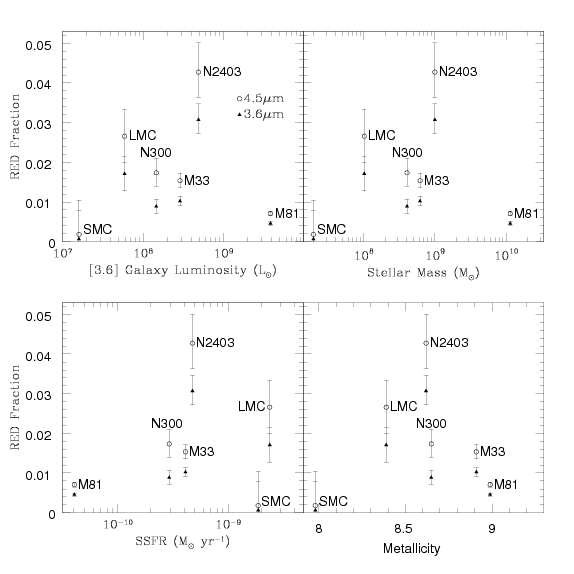}
\figurenum{6}
\label{fig:redplot}
\caption{The fraction of the 3.6\micron~(filled triangles) and 4.5\micron~(open circles) luminosity produced by stars in the red star region as a function of galaxy luminosity (top left), stellar mass (top right), specific star formation rate (lower left) and metallicity (lower right). }
\end{figure}

The ``red stars'' region is a mixture of sources.  Most should be massive, evolved stars such as $\eta$ Carinae \citep{humphreys1994}, IRC+10420 (e.g. \citealt{jones1993}), or M33 Object X \citep{khan2011} which are obscured by dust formed in mass ejections.  Some are young star clusters containing dust, such as the brightest 3.6~$\micron$ source in M33 examined by \citet{khan2011} in the process of finding Object X.  Explosive transients can also form dust and lie in this region (\citealt{fox2011}; \citealt{dorota2012}).
Since all these sources are likely massive stars or young clusters, we expected that the emissions from these red objects would be tied to the SFR.  As shown in \ref{fig:redplot}, We see no correlation in our full sample between the bright, red source luminosity contribution and any galaxy property, including the SFR or SSFR.
However, if we ignore the SMC, then there is a relatively clean trend with SSFR.  \citet{bonanos2010} examined optically selected luminous stars in the LMC and SMC, finding that those in the SMC seemed to be less dusty. \citet{boyer2011} found similar trends. 
Perhaps what we are observing is a combination of metallicity and SSFR, with the numbers rising with SFR but with dust production truncated at the lowest metallicity. 
 This causes the bright, red stars of the SMC to be less red and dimmer in these IR bands. 
On average, the contribution of these sources is small, with $1.2\%$ of the 3.6~$\micron$ flux and 1.9\% of the 4.5~$\micron$ flux due to these bright, but rare obscured sources. 

To put the impact of the IR light from these dusty stars into the broader context of the galaxy as a whole, we can compare our measurements to the bolometric luminosities of the galaxies.  
Using the galaxy SED templates of \citet{assef2010} to estimate the bolometric luminosity of the galaxies, the summed 3.6 and 4.5~$\micron$ emission represents $\sim$ 5\% of the ``bolometric'' luminosity from 0.1-30~$\micron$, so the summed 3.6 and 4.5~$\micron$ emission of DAGB stars ranges from $\sim$ 0.04\% (M81) to $\sim$ 0.37\% (SMC) of the bolometric luminosity.
If we model the mid-IR fluxes of the DAGB stars as a stellar black body surrounded by a dusty shell using DUSTY \citep{ivezic1999}, we can estimate their total mid-IR luminosity.  We find preferred dust temperatures of 1000-1500~K, as expected for emissions from dusty AGB winds with significant visual and negligible mid-infrared optical depths.  These simple estimates are not model dependent. 
In these models, the summed 3.6 and $4.5~\micron$ DAGB luminosity represents $\sim50\%$ of the hot dust emission from these stars.  This means that the total dust emission of the DAGB stars represents of order $\sim 0.1\%$ (M81) to $\sim 0.8\%$ (SMC) of the total luminosity of the galaxies.
 The difference between the observed 3.6 and $4.5~\micron$ emission and the total is modest because the dust is relatively hot.  We cannot accurately estimate a similar correction for the red sources because they likely peak at a longer wavelengths ($\sim$24~$\micron$) and we cannot effectively estimate a dust temperature.
Overall, we find that circumstellar dust around evolved stars represents a modest correction for galaxy SEDs but can be a significant correction to their mid-IR colors.

\section{Discussion}
We surveyed the luminous dusty stars in 6 nearby galaxies. Using 3.6 and 4.5~$\micron$ data we identified DAGB and other luminous red sources with significant emission from hot circumstellar dust.  We determine their contribution to the flux at these bands and whether there is any correlation between their luminosity contribution and galaxy metallicity, stellar mass, luminosity or SSFR.  Such studies can help to constrain one of the uncertainties considered in \citet{conroy2010}, namely the fraction of stellar emission obscured by stellar dust.  By selecting the stars in the mid-IR we automatically include heavily obscured sources that would be missed in optical studies.  Additionally, by looking at whole galaxies, we have a relatively large sample of these relatively rare stars and should obtain reliable population statistics.  

The contributions of obscured DAGB stars range from 0.9\% (M81) to 5.6\% (SMC) at $3.6~\micron$ and from 1.0\% (M81) to 10.1\% (SMC) at 4.5~$\micron$. 
We see a trend of higher mid-IR with metallicity, SSFR, stellar mass and luminosity, but this is most likely a correlation with metallicity since metallicity is correlated with the other properties and none of the other properties should be correlated with DAGB stars.  It is a legitimate concern, however, that without complete star formation histories, we cannot be certain the SFR does not significantly effect the correlation.

The dependence of AGB dust production and the resulting IR emission on initial metallicity is an area of active study, with somewhat conflicting results.  It is a difficult problem, combining stellar evolution, chemistry, wind formation and dust properties. 
For example, \citet{ventura2012} examined models of dust production by AGB stars at LMC and SMC metallicities.  They find that for the more massive stars that undergo hot bottom burning ($8M_{\odot} > M > 3.5M_{\odot}$), a higher metallicity results in more dust production. For lower mass stars that undergo third dredge up ($3.5M_{\odot} > M > 1.5M_{\odot}$), they find that the dust production is nearly independent of metallicity.  The more massive stars become oxygen rich at the surface, in which case dust formation is limited by the availability of silicon, which is controlled by the metallicity. 
For the low mass stars, third dredge up leads to an enhanced surface abundance of carbon and thus dust formation that depends little on initial metallicity.  \citet{ventura2012} in the end argue that theoretical predictions of dust production around lower mass AGB stars are not robust due to uncertainties in the amount of mixing and the extent of the third dredge up.  
Similar results are found by \citet{marigo2008}, \citet{bowen1991} and \citet{matsuura2007}.
\citet{groenewegen2009} compare their sample of LMC and SMC AGB stars to models for mass loss, taking different dust grain composition into account, and compare to Galactic estimates to find that there is no strong dependence of mass-loss rate on metallicity to factors of order 2-4.  This is broadly consistent with \citet{wachter2008}, who find the mass loss rate of the Milky Way AGB stars to be a factor of 2 higher than that of the SMC.  
 While higher metallicity may result in more dust formation, this dust is dominated by silicates which have a lower opacity than graphitic dust.  These issues make it difficult to translate any trends into observational predictions. 

The \citet{maraston2005} models of the AGB contribution are based on the fuel consumption theorem \citep{renzini1981} and assume mass loss is not significantly effected by metallicity.  Instead, it is the surface abundances that are modified, so that the number ratio of oxygen-rich to carbon-rich AGB stars depends on metallicity. A metal-poor population is expected to have more carbon-rich stars.  
 A metal-poor AGB star has a lower abundance of oxygen in its envelope, so less carbon has to be dredged up before the oxygen has all been bound in CO and the residual carbon can go on to form dust (\citealt{iben1983}; \citealt{maraston2006}).  In this scenario, a metal-poor population could result in a higher mid-IR luminosity.  
\citet{bird2011} examined the energy output of AGB stars using fuel consumption theory, but find insufficient calibration data to determine an accurate metallicity dependence. 
The earlier Padova models (\citealt{marigo2007}; \citealt{marigo2008}) have more carbon-rich stars at low metallicity due to the higher efficiency of third dredge up, while the more recent \citet{girardi2010} models have increased mass loss rates and shorter AGB phase lifetimes at low metallicity.  While the effect of metallicity is not fully explored by these studies, the increased dust production and mass loss rate suggested at low metallicity would result in a higher mid-IR contribution, as we find in our sample. 

Previous observational constraints on the metallicity dependence of the dust production also have mixed conclusions. 
\citet{meidt2012b} found that lower metallicity clusters in M~100 have more dusty AGB stars, arguing that since the dust optical depth depends upon whether an AGB star is O-rich or C-rich, the AGB contribution to the cluster depends upon metallicity.  They also find larger mass-loss rates for younger clusters.  Since the younger clusters also have a higher metallicity, they cannot address the overall effect of metallicity on the AGB contribution. 
Looking at the AGB population of eight local group dwarf irregular galaxies, \citet{boyer2009} find that all galaxies in their sample have obscured AGB stars.  For AGB stars with optical counterparts, they find that the more metal-rich galaxies have a larger population of red, dust enshrouded AGB stars.  However, requiring optical counterparts biases the sample against dustier stars.  
An alternate explanation for the increased number of dusty stars in the more metal-rich galaxies is the age of the populations, because the higher metallicity galaxies in their sample also have more recent ($<$ 3 Gyr) star formation. 
They find no clear correlation between metallicity and optical completeness, as would be expected if metallicity were the dominating factor in the degree of obscuration.

\citet{meidt2012a} examined the contribution from red and AGB stars to the luminosity of M81.
 They find hot dust and polycyclic aromatic hydrocarbons (PAHs) contributes 5.0$\pm0.9\%$ while red and AGB stars contribution only 0.10$\pm0.02\%$ of the $3.6~\micron$ luminosity, rather than our significantly higher estimate of 0.58$\pm0.02\%$ for the DAGB stars.
  \citet{meidt2012a} used a very different method to estimate the percentage of the light due to AGB and intermediate age stars.
 Essentially, they divide the 3.6 and 4.5~$\micron$ emission into that from stars and that due to ``contaminating'' sources that contribute to the mid-IR flux while representing little stellar mass.  This includes, dusty stars of all luminosities and PAH emission.  They do this by using independent component analysis (ICA) to separate out statistically significant source distributions and maximize $[3.6]-[4.5]$ color differences.  Using the $8.0~\micron$ images, they were able to obtain a rough estimate of the contribution of stellar and  non-stellar (dust) sources. They note a selection bias against spatially coincident dust and clusters dominated by intermediate age stars and also a possible bias from mismatched [3.6] and [8.0] PSFs which would confuse the division between stellar and non-stellar emission.  They caution the distinction between 3.6~$\micron$ emission due to evolved stars and PAH emission are good for ``rough estimates'' of the components and that a more detailed analysis of the ``contaminates'' is possible but beyond their scope.  However, \citet{meidt2012b} explores evolved stars in M100 in more detail.

For comparison, \citet{melbourne2012} estimate the near-IR (HST WFC3/IR F160W) H-band contributions of AGB stars in three of our galaxies, M81, NGC~300 and NGC~2403.  The estimates are only for a single WFC3 fields rather than global estimates.  They find a NIR AGB luminosity contribution of 8\% for M81, 15\% for NGC~300 and 17\% for NGC~2403.  Since the $1.6~\micron$ emission is dominated by the photosphere of the AGB stars rather than the reprocessed emission from dust around stars, these fractions need not match ours. 
Like \citet{kriek2010}, they find that the stellar population models over-predict the AGB NIR luminosity and that the $1.6~\micron$ AGB luminosity fraction increases with the mass fraction of intermediate age stars ( $<$ 2 Gyr and $<$ 0.3 Gyr) with no obvious metallicity trends.

The brighter, red sources contribute less, typically accounting for $\sim$1.2\% of the $3.6~\micron$ and $\sim$1.8\% of the $4.5~\micron$ luminosity.
These sources are a mixture of dusty young massive stars and star clusters, and we expect their luminosity fraction could be strongly correlated with the SSFR.  The correlation may exist if there is also strong suppression of the red stars at the low metallicity of the SMC (\citealt{bonanos2009}; \citealt{bonanos2010}).  We lack a large enough distribution of galaxy properties to address this question. 
These sources are probably more important at longer wavelengths.  The 3.6~$\micron$ and 4.5~$\micron$ emissions are due to fairly hot dust close to the stars, while many of these sources are episodic dust sources rather than having long lived winds (see discussion in \citet{kochanek2012}).  During most of the period where they have a significant dust optical depth, the dust is at larger distances and cooler temperatures radiating at longer wavelengths.  Simple estimates suggest that their contribution will peak $\sim$24~$\micron$, similar to $\eta$ Carinae's present day SED (\citealt{robinson1973}; \citealt{humphreys1994}).  Dusty young star clusters will also tend to be dominated by cooler dust \citep{whelan2011}. 

Overall, our findings are in agreement with other recent studies, finding that while (D)AGB stars have a significant effect on a galaxy's SED in the IR, the contribution is not as large as originally thought (\citealt{kelson2010}; \citealt{boyer2011}; \citealt{meidt2012b}).
 \citet{boyer2011} find that AGB stars contribute about 20\% of the 3.6~$\micron$ flux in the SMC, compared to our $\sim$6\%. Our mid-IR identification of sources, likely selects a smaller, dustier sample of stars causing this difference. 

We also examined how the dusty stars we study effect the SED of a galaxy.  By comparing to galaxy templates from \citet{assef2010}, we found that the summed 3.6 and 4.5~$\micron$ emission represents $\sim$ 5\% of a galaxy's bolometric luminosity, and that the DAGB star contribution ranges from $\sim$ 0.04\% (M81) to $\sim$ 0.37\% (SMC) of the bolometric luminosity.  By modeling the AGB star contribution with DUSTY \citep{ivezic1999} we found that the DAGB luminosity constitutes $\sim$ 50\% of the hot dust emission from these stars.  This means that while dusty stars have a small effect on the overall galaxy SED, they can significantly effect the mid-IR colors.
Note that here we have focused on the contribution of dust emission by AGB stars to the overall luminosity of the galaxy, while many of these other studies are focused on the contribution from the stellar photospheres of the AGB stars. 
These less or non-dusty AGB stars likely contribute a significant amount of mid-IR flux without distorting the SED enough to produce a red mid-IR color. 

 The uncertainty in the treatment of the AGB phase in the SPS models can have a significant impact on the estimation of galaxy properties (\citealt{maraston2006}; \citealt{conroy2009}; \citealt{conroy2010}; \citealt{conroy2010b}).  
For example, \citet{conroy2010} find that lower metallicity populations prefer cooler AGB stars which could in part reflect increased dust production at lower metallicity.  They also find that both TP-AGB stars and dust cause significant uncertainties in the properties of star forming galaxies.  
In studies like \citet{conroy2010}, it would be straight forward to include a reasonable phenomenological model of AGB or other stars with dusty winds.  For a period $T_{wind}$ stars are assigned a dust optical depth $\tau$ with the dust radiating at $T_{dust}\sim1000$ K.  The SED can easily be calculated using DUSTY \citep{ivezic1999} or similar dust radiation transfer models.  Extending the UV through near-IR wavelengths of \citet{conroy2010} and \citet{conroy2010b} to the mid-IR would then constrain the hot dust contribution and its effects on the overall SED because it will be difficult for other variables to mimic the mid-IR dust emission.  Focusing on the 4.5 and 5.8 $\micron$ bands would minimize the need to worry about the strong PAH emissions at 8 $\micron$ and the weaker emissions at 3.6 $\micron$.
 Although the overall amount of flux from dusty stars may be small, the effect on the mid-IR can be substantial and should be explored further. 
Our results are primarily limited by the low resolution of Spitzer.  The James Webb Space Telescope will allow this type of study to be performed on many more galaxies and over a broader wavelength range, providing better constraints on how hot dust around stars, and AGB stars in particular, effect a galaxy's SED and how their contribution correlates with galaxy properties such as SFH and metallicity.

\acknowledgments 

Acknowledgments: 
The authors thank R. Khan, K.Z. Stanek, M.H. Pinsonneault and T.A. Thompson for their helpful discussions and comments.
We also thank the SINGS Legacy and LVL Surveys for making their data publicly available.
This work is based in part on observations made with the Spitzer Space Telescope, which is operated by the Jet Propulsion Laboratory, California Institute of Technology under a contract with NASA.
J.R.G. and C.S.K. are supported by NSF grant AST-0908816.

\begin{deluxetable}{ccccccc}
\tablecolumns{7}
\tablewidth{0pc}  
\tablecaption{ \label{tab:prop} Galaxy Properties} 
\tablehead{
  \colhead{ID} & \colhead{Distance} &  \colhead{Metallicity} & \colhead{Stellar Mass} &  \colhead{SFR} & \colhead{SSFR}\\
              & \colhead{Mpc} &  \colhead{12+log(O/H)} & \colhead{log($M_{*}/M_{\odot}$)} & \colhead{$M_{\odot}$yr$^{-1}$} & \colhead{Gyr$^{-1}$} }
\startdata
M33      & 0.96 & 8.91 & ~8.79 & 0.25 &  0.41  \\
M81      & 3.65 & 8.99 & 10.06 & 0.46 &  0.04  \\
NGC~300  & 2.00 & 8.65 & ~8.61 & 0.12 &  0.29  \\
NGC~2403 & 3.22 & 8.62 & ~9.00 & 0.47 &  0.47  \\
SMC      & 0.06 & 7.98 & ~7.29 & 0.04 &  1.54  \\
LMC      & 0.05 & 8.39 & ~8.00 & 0.24 &  1.96  \\ 
\enddata

\tablecomments{The distances are from \citet{zaritsky1994}, expect for M33 \citep{bonanos2006} and M81 \citep{gerke2011}.  The metallicities are from \citet{zaritsky1994} except for the SMC \citep{peimbert1976} and LMC \citep{pagel1978}.  The stellar masses were determined using K band luminosities from \citet{dale2009} and the mass to light ratio from \citet{bell2003}.  For the SMC and LMC the stellar masses were determined using the \citet{dale2009} [3.6] magnitude and an averaged mass to light ratio from \citet{falcon2011}.  The star formation rates were calculated using H$\alpha$ measurements from \citet{kennicutt2008} with the conversion from \citet{meurer2009}.
}
\end{deluxetable}

\begin{deluxetable}{ccccc}
\tablecolumns{5}
\tablewidth{0pc}  
\tablecaption{ \label{tab:totL} Total Galaxy Luminosity} 
\tablehead{
  \colhead{ID} & \multicolumn{2}{c}{Raw (L$_{\odot}$)}& \multicolumn{2}{c}{Corrected (L$_{\odot}$)} \\
   & \colhead{[3.6]} & \colhead{[4.5]} & \colhead{[3.6]} & \colhead{[4.5]}
}
\startdata
M33      & 3.78$\times 10^{8}$  & 2.70$\times 10^{8}$ & $(2.88\pm0.01)\times 10^{8}$ & $(1.49\pm0.004)\times 10^{8}$ \\
M81      & 3.93$\times 10^{9}$ & 1.93$\times 10^{9}$ & $(3.84\pm0.01)\times 10^{9}$ & $(1.86\pm0.003)\times 10^{9}$ \\
NGC~300  & 1.60$\times 10^{8}$  & 8.80$\times 10^{7}$ & $(1.47\pm0.09)\times 10^{8}$ & $(7.60\pm0.502)\times 10^{7}$ \\
NGC~2403 & 5.59$\times 10^{8}$  & 3.16$\times 10^{8}$ & $(4.89\pm0.09)\times 10^{8}$ & $(2.61\pm0.041)\times 10^{8}$ \\
SMC      & 2.81$\times 10^{7}$  & 1.49$\times 10^{7}$ & $(1.61\pm0.17)\times 10^{7}$ & $(8.09\pm1.124)\times 10^{6}$ \\
LMC      & 1.11$\times 10^{8}$  & 5.69$\times 10^{7}$ & $(5.91\pm0.23)\times 10^{7}$ & $(2.98\pm0.131)\times 10^{7}$ \\ 

\enddata
\end{deluxetable}

\begin{deluxetable}{cc}
\tablecolumns{2}
\tablewidth{0pc}  
\tablecaption{ \label{tab:cmd} CMD Limits} 
\tablehead{
  \colhead{Region} & \colhead{Limits}
}
\startdata
MS slant: & $M_{4.5} < 4.375([3.6]-[4.5])-10.563$\\
MS line: & $[3.6]-[4.5] < 0.10$ \\
blue SM limit: & $[3.6]-[4.5] > -0.75$ \\
MS upper limit: & $M_{4.5} > -13.60$\\
MS lower limit: & $M_{4.5} < -8.75$\\

red upper limits: & $M_{4.5} > -15.50 $\\
red upper slant: & $M_{4.5} > -2.727([3.6]-[4.5])-13.327$\\

AGB bottom slant: & $M_{4.5} < -1.271([3.6]-[4.5])-8.229$\\
AGB bottom limit: & $M_{4.5} < -9.50 $\\
AGB top slant: & $M_{4.5} > -3.437([3.6]-[4.5])-9.781$\\
AGB top limit: & $M_{4.5} > -11.50 $\\
\enddata
\end{deluxetable}

\begin{deluxetable}{ccccccc}
\tablecolumns{7}
\tablewidth{0pc}  
\tablecaption{ \label{tab:bg} Background Contamination} 
\tablehead{
  \colhead{ID} & \multicolumn{2}{c}{AGB}& \multicolumn{2}{c}{Red}& \multicolumn{2}{c}{Blue}\\
 & \colhead{[3.6]} &  \colhead{[4.5]} & \colhead{[3.6]} & \colhead{[4.5]} & \colhead{[3.6]} & \colhead{[4.5]}}
\startdata
M33 &       11 & 11 & ~3 & ~2 & 28 & 31 \\
M81 &       31 & 31 & 28 & 26 & 30 & 32 \\
NGC~300 &   20 & 19 & 11 & ~7 & 48 & 47 \\
NGC~2403 &  43 & 42 & 29 & 31 & 31 & 30 \\
SMC  &      19 & 18 & 93 & 84 & 61 & 61 \\
LMC  &      32 & 32 & 21 & 17 & 62 & 62 \\

\enddata
\tablecomments{These are the luminosities in the background regions as a percentage of the luminosity of the galaxy after correcting for area.}
\end{deluxetable}

\begin{deluxetable}{ccccccc}
\tablecolumns{7}
\tablewidth{0pc}  
\tablecaption{ \label{tab:fracR} Raw Luminosity Fractions} 
\tablehead{
  \colhead{ID} & \multicolumn{2}{c}{AGB region} & \multicolumn{2}{c}{Red stars}& \multicolumn{2}{c}{Blue stars}\\
        & \colhead{[3.6]} &  \colhead{[4.5]} & \colhead{[3.6]} &  \colhead{[4.5]} & \colhead{[3.6]} &  \colhead{[4.5]} }
\startdata

M33      &  $4.29\pm{0.11}$  &  $~7.37\pm{0.19}$  &  $1.42\pm{0.16}$  &  $2.09\pm{0.24}$  &  $14.96\pm{0.39}$  &  $13.50\pm0.36$ \\
M81      &  $0.86\pm{0.01}$  &  $~1.44\pm{0.02}$  &  $0.65\pm{0.03}$  &  $0.98\pm{0.05}$  &  $~2.12\pm{0.04}$  &  $~2.06\pm0.04$ \\
NGC~300  &  $3.59\pm{0.12}$  &  $~6.29\pm{0.21}$  &  $1.06\pm{0.17}$  &  $1.98\pm{0.35}$  &  $~7.78\pm{0.35}$  &  $~7.31\pm0.33$ \\
NGC~2403 &  $4.27\pm{0.08}$  &  $~6.88\pm{0.13}$  &  $4.45\pm{0.22}$  &  $6.20\pm{0.39}$  &  $~8.59\pm{0.27}$  &  $~7.78\pm0.25$ \\
SMC      &  $7.49\pm{0.57}$  &  $13.66\pm{1.10}$  &  $1.00\pm{0.47}$  &  $1.26\pm{0.63}$  &  $55.35\pm{3.39}$  &  $53.90\pm3.56$ \\
LMC      &  $7.44\pm{0.28}$  &  $13.86\pm{0.53}$  &  $2.13\pm{0.30}$  &  $3.24\pm{0.47}$  &  $44.09\pm{1.24}$  &  $41.24\pm1.19$ \\

\enddata
\tablecomments{The fractions are expressed as a percent of the total 3.6 or 4.5\micron~luminosities from Table \ref{tab:totL}.}
\end{deluxetable}

\begin{deluxetable}{cccccccc}
\tablecolumns{8}
\tablewidth{0pc}  
\tablecaption{ \label{tab:fracC} Background-Corrected Luminosity Fractions } 
\tablehead{
  \colhead{ID} & \multicolumn{2}{c}{AGB} & \multicolumn{2}{c}{Red}& \multicolumn{2}{c}{Blue}\\
        & \colhead{[3.6]} &  \colhead{[4.5]} & \colhead{[3.6]} &  \colhead{[4.5]} & \colhead{[3.6]} &  \colhead{[4.5]} }
\startdata

M33     &  $2.85\pm0.10$ & $~4.94\pm0.17$  & $1.02\pm0.12$ & $1.54\pm0.18$ & $~8.00\pm0.72$ & $~7.02\pm0.72$ \\
M81     &  $0.58\pm0.02$ & $~0.97\pm0.03$  & $0.46\pm0.04$ & $0.71\pm0.06$ & $~1.47\pm0.09$ & $~1.37\pm0.09$ \\
NGC~300 &  $2.71\pm0.15$ & $~4.83\pm0.26$  & $0.89\pm0.18$ & $1.73\pm0.35$ & $~3.80\pm1.00$ & $~3.67\pm0.89$ \\
NGC~2403&  $2.38\pm0.14$ & $~3.98\pm0.23$  & $3.07\pm0.38$ & $4.27\pm0.69$ & $~5.81\pm0.54$ & $~5.40\pm0.46$ \\
SMC     &  $5.55\pm0.86$ & $10.12\pm1.81$  & $0.06\pm0.67$ & $0.18\pm0.80$ & $19.61\pm6.96$ & $19.06\pm6.84$ \\
LMC     &  $5.03\pm0.52$ & $~9.47\pm0.98$  & $1.71\pm0.44$ & $2.66\pm0.67$ & $16.63\pm3.35$ & $15.65\pm3.07$ \\
\enddata
\tablecomments{The fractions are expressed as a percent of the total 3.6 or 4.5\micron~luminosities from Table \ref{tab:totL}.}
\end{deluxetable}


\end{document}